\documentclass[aps, prd,  superscriptaddress, showpacs, longbibliography, floatfix, nofootinbib,10pt]{revtex4-2}

\usepackage{bm, amsmath, amssymb, relsize, amsfonts, mathrsfs, multirow, braket, siunitx, color, booktabs, arydshln, tabularx}
\usepackage{graphicx}
\graphicspath{{./figures/}}

\DeclareSIUnit \parsec {pc}

\usepackage[dvipsnames]{xcolor}
\usepackage[unicode]{hyperref}
\hypersetup{
  colorlinks=true,
  citecolor=RoyalBlue,
  linkcolor=blue,
  urlcolor=blue
}

\usepackage{array}
\usepackage{cancel}
\usepackage{orcidlink}
\usepackage{graphicx}
\usepackage{dcolumn}
\usepackage{bm}
\usepackage{paralist}
\usepackage{epsfig}
\usepackage{placeins}
\usepackage{mathrsfs}
\usepackage{comment}
\usepackage{color, colortbl}
\usepackage[inline]{enumitem}
\usepackage{soul}
\usepackage{adjustbox}
\usepackage{amssymb}
\usepackage{caption}
\usepackage{subcaption}
\usepackage{float}
\usepackage{cancel}
\usepackage{verbatim}
\usepackage{amsmath}
\usepackage[toc,page]{appendix}
\usepackage{graphicx}
\DeclareUnicodeCharacter{2212}{\ensuremath{-}}

\DeclareMathAlphabet{\mathpzc}{OT1}{pzc}{m}{it}
	
\definecolor{LightCyan}{rgb}{0.88,1,1}
\definecolor{lightgray}{gray}{0.9}

\def \IITGn     {Department of Physics, Indian Institute of Technology Gandhinagar, Gujarat 382055, India.\vspace*{4pt}}
\def \HRI     {Regional Centre for Accelerator-based Particle Physics, Harish-Chandra Research Institute, HBNI, Chhatnag Road, Jhunsi, Prayagraj (Allahabad) 211019, India\vspace*{4pt}}
\def \UoS     {School of Physics \& Astronomy, University of Southampton, Highfield, Southampton SO17 1BJ, UK.\vspace*{4pt}}
\def \UU     {Department of Physics \& Astronomy, Uppsala University, Box 516, 75120 Uppsala, Sweden.\vspace*{4pt}}

\begin{document}

\title{Explaining 95 GeV Anomalies 
in the 2-Higgs Doublet Model Type-I}

\author{\textsc{Akshat Khanna}\orcidlink{0000-0002-2322-5929}\vspace*{7pt}}
\email{khanna$_$akshat@iitgn.ac.in}
\affiliation{\IITGn}

\author{\textsc{Stefano Moretti}\orcidlink{0000-0002-8601-7246}\vspace*{7pt}}
\email{s.moretti@soton.ac.uk; stefano.moretti@physics.uu.se}
\affiliation{\UoS}
\affiliation{\UU}

\author{\textsc{Agnivo Sarkar}\orcidlink{0000-0001-9596-1936}}
\email{agnivosarkar@hri.res.in}
\affiliation{\HRI}

\begin{flushright}
    HRI-RECAPP-2024-05
\end{flushright}

\begin{abstract}
We show how the 2-Higgs Doublet Model (2HDM)  Type-I can explain some excesses recently seen at the Large Hadron Collider (LHC) in $\gamma\gamma$ and $\tau^+\tau^-$ final states in turn matching Large Electron Positron (LEP) data in $b\bar b$ signatures, all anomalies residing around 95 GeV. The explanation to such anomalous data is found in the aforementioned scenario when in inverted mass hierarchy, in two configurations: i) when the lightest CP-even Higgs state is alone capable of reproducing the excesses; ii) when a combination of such a state and the CP-odd Higgs boson is able to do so. To test further this scenario, we present some Benchmark Points (BPs)  of it amenable to phenomenological investigation.   
\end{abstract}
\maketitle

\section{Introduction}
\label{sec:intro}

A long-standing anomaly existing in LEP collider data \cite{LEPWorkingGroupforHiggsbosonsearches:2003ing} is the one hinting at the possibility of  $e^+e^-\to Zh$ events being produced therein, with a Higgs boson state $h$ with a mass of approximately 98 GeV decaying into ${b\bar b}$ pairs \cite{ALEPH:2006tnd}. More recently, the CMS collaboration at the LHC has  found an excess near 95 GeV in di-photon events in two separate analyses  \cite{CMS:2018cyk,CMS:2023yay}.
In fact, they also reported an excess in $\tau^+\tau^-$ pairs, again, around a mass of about 98 GeV. Finally, ATLAS also observed an excess at around 95 GeV in di-photon events, thereby aligning with  CMS, although,  especially when including `look elsewhere' effects, their findings are far less significant than the CMS ones.  Altogether, in view of the limited mass resolution of the di-jet invariant mass at LEP, this older anomaly may well be consistent with the  excesses seen by CMS (and, partially) ATLAS in the $\gamma\gamma$ and, even more so, $\tau^+\tau^-$ final states (as the mass resolution herein is also rather poor).
As a consequence of this credible mass overlap, many  studies \cite{Cao:2016uwt,Heinemeyer:2021msz,Biekotter:2021qbc,Biekotter:2019kde,Cao:2019ofo,Biekotter:2022abc,Iguro:2022dok,Li:2022etb,Cline:2019okt,Biekotter:2021ovi,Crivellin:2017upt,Cacciapaglia:2016tlr,Abdelalim:2020xfk,Biekotter:2022jyr,Biekotter:2023jld,Azevedo:2023zkg,Biekotter:2023oen,Cao:2024axg,Wang:2024bkg,li2023lightdarkmatterconfronted,Dev:2023kzu,Borah:2023hqw,Cao:2023gkc,Ellwanger:2023zjc,Aguilar-Saavedra:2023tql,Ashanujjaman:2023etj,Dutta:2023cig,Ellwanger:2024txc,Diaz:2024yfu,Ellwanger:2024vvs,YaserAyazi:2024hpj, Ahriche_2024, bhattacharya2023growingexcessesnewscalars} have tested the possibility of simultaneously fitting these excesses within Beyond the Standard Model (BSM) frameworks featuring a non-SM Higgs state lighter than 125 GeV, i.e., than the one observed at the LHC in 2012 \cite{ATLAS:2012yve,CMS:2012qbp}.

A possible route to follow in explaining such events through a companion Higgs state (to the 125 GeV one) is to resort to a  2HDM \cite{Gunion:1992hs,Branco:2011iw}, as done, e.g., in Refs. \cite{Benbrik:2022azi, Benbrik:2022tlg, Belyaev:2023xnv,Benbrik:2024ptw,Benbrik:2024ptw}, wherein a Type-III (which allows direct couplings of both Higgs doublets to all SM fermions) with specific fermion textures was invoked successfully as a BSM explanation to the $b\bar b$, $\gamma\gamma$ and $\tau^+\tau^-$ excesses seen at LEP and the LHC. Herein, 
both a fully CP-even and a mixed CP-even/odd solution was found, upon refining the 2HDM Type-III Yukawa structure to comply with both  theoretical consistency requirements and experimental measurements of the discovered Higgs mass and couplings (of the 125 GeV Higgs state).

In this study, we would like to analyse whether solutions of the same kind (i.e., both a fully CP-even and a mixed CP-even/odd one) can also be found in a 2HDM Type-I, wherein only one Higgs doublet gives mass to all SM fermions, again, satisfying the aforementioned  theoretical requirements and experimental constraints.

The  paper is  organised as follows. In the next section, we briefly review the theoretical framework of the 2HDM Type-I. Then we discuss the theoretical and experimental constraints applied to such a BSM scenario in our study, which are described in section~\ref{sec:constraints}. 
In section~\ref{sec:anomaly} we show how the aforementioned configurations can naturally explain the discussed anomalous data. We then summarise our findings in section~\ref{sec:conclusions}. 

%after which we move on to show how the latter can naturally explain the discussed anomalous data in various configurations of its parameters space. Our conclusions  then follow.  

\section{The 2HDM Type-I}
\label{sec:model}
Among the various BSM scenarios, the 2HDM can be considered as a simple extension of the SM. In fact, the (pseudo)scalar sector of this model comprises two complex Higgs  fields, $\phi_{1}$ and $\phi_{2}$, which transform as a doublet under the  Electro-Weak (EW) gauge group $SU(2)_{L}\times U(1)_{Y}$ with hypercharge $Y = 1$. (For a detailed overview of this model interested reader can look into ~\cite{Branco_2012}.) The most general gauge invariant CP-conserving scalar potential can be written as 
\begin{equation}
    \label{eq:fullpot}
     \begin{split}
    V(\phi_{1}, \phi_{2}) & = m_{11}^2 \phi_1^\dagger \phi_1 + m_{22}^2 \phi_2^\dagger \phi_2 - m^{2}_{12}\left[\phi^{\dagger}_{1}\phi_{2} + \phi^{\dagger}_{2}\phi_{1}\right] + \frac{\lambda_1}{2}(\phi_1^\dagger \phi_1)^2 + \frac{\lambda_2}{2}(\phi_2^\dagger \phi_2)^2 \\
    & + \lambda_3(\phi_1^\dagger \phi_1)(\phi_2^\dagger \phi_2) + \lambda_4(\phi_1^\dagger \phi_2)(\phi_2^\dagger \phi_1) + \left[  \frac{\lambda_5}{2}(\phi_1^\dagger \phi_2)^2 + h.c. \right] + \{ \left[ \lambda_{6}(\phi_1^\dagger \phi_1
    )+ \lambda_{7}(\phi_2^\dagger \phi_2) \right](\phi^{\dagger}_{1}\phi_{2}) + h.c.  \}.
    \end{split}
\end{equation}
Given the hermiticity of the scalar potential, all the potential parameters of Eq.~(\ref{eq:fullpot}) must be real. In order to prevent tree level Flavour Changing Neutral Currents (FCNCs), one can postulate an additional discrete $\mathcal{Z}_{2}$ symmetry under which the scalar fields transform as $\phi_{1} \to \phi_{1}$, $\phi_{2} \to -\phi_{2}$. One then realises that the term proportional to $m^{2}_{12}$, $\lambda_{6}$ and $\lambda_{7}$ in Eq.~(\ref{eq:fullpot}) violates this $\mathcal{Z}_{2}$ symmetry explicitly. Therefore the potential can be expressed in the following form:
\begin{equation}
    \label{eq:scalarpot}
    \begin{split}
    V(\phi_1, \phi_2) & = m_{11}^2 \phi_1^\dagger \phi_1 + m_{22}^2 \phi_2^\dagger \phi_2 + \frac{\lambda_1}{2}(\phi_1^\dagger \phi_1)^2 + \frac{\lambda_2}{2}(\phi_2^\dagger \phi_2)^2 \\
    & + \lambda_3(\phi_1^\dagger \phi_1)(\phi_2^\dagger \phi_2) + \lambda_4(\phi_1^\dagger \phi_2)(\phi_2^\dagger \phi_1) + \{   \frac{\lambda_5}{2}(\phi_1^\dagger \phi_2)^2 + h.c. \}.
    \end{split}
\end{equation}

\noindent
Both these Higgs fields, $\phi_1$ and $\phi_2$, acquire a non-zero Vacuum Expectation Value
({VEV}) (i.e., $\langle\phi_{1}\rangle = v_{1}$ and $\langle\phi_{2}\rangle = v_{2}$) and spontaneously break the EW gauge symmetry down to $U(1)_{\rm EM}$. After symmetry breaking, the $W^\pm$ and $Z$ boson becomes massive and the (pseudo)scalar sector contains five physical Higgs bosons - two CP-even $\{H, h\}$, one CP-odd $A$ and a pair of charged states $H^{\pm}$ with masses $m_{H}, m_{h}, m_{A}$  and $m_{H^{\pm}}$,  respectively. Both these VEVs $v_1$ and $v_2$ are related to the EW scale $v = \sqrt{v^{2}_{1} + v^{2}_{2}}$ = 246 GeV. The ratio between these two VEVs can be parameterised as $\tan\beta = \frac{v_2}{v_1}$. In addition, the mixing angle between the CP-even states $\{H, h\}$ can be parametrised as $\alpha$. For the present study we consider the inverted hierarchy between the CP-even mass eigenstates. This particular choice alters the usual interpretation of the mass spectrum and the couplings\footnote{For example, in the inverted mass hierarchy the alignment limit corresponds to $\cos(\beta - \alpha) \to 1$. However, we will not conform strictly to this limit in the present study.}. Hereafter we align $H$ as to be the SM like Higgs boson and $h$ to be the lighter scalar state. In Eq. (\ref{Eq:lambda}) we present the relations between the potential parameters $\lambda_{i}$'s with the physical parameters of the model:  

\begin{equation}\label{Eq:lambda}
    \begin{split}
    \lambda_1 & = \frac{c_\alpha^2m_{H}^2 + s_\alpha^2m_{h}^2}{v^2c_{\beta}^2}, \\
    \lambda_2 & = \frac{c_\alpha^2m_{h}^2 + s_\alpha^2m_{H}^2}{v^2s_{\beta}^2}, \\
    \lambda_3 & = \frac{(m_{H}^2-m_{h}^2)s_{\alpha}c_{\alpha} - (\lambda_4+\lambda_5)v^2c_{\beta}s_{\beta}}{v^2c_{\beta}s_{\beta}}, \\
    \lambda_4 & = \frac{m_A^2 - 2m_{H^\pm}^2}{v^2}, \\
    \lambda_5 & = -\frac{m_A^2}{v^2}.
    \end{split}
\end{equation}
The scalar potential given in Eq.~(\ref{eq:scalarpot}) clearly has six independent parameters, so that the $m_{ii}^2$'s can be traded off in terms of $\lambda_i$'s using the extremisation conditions on the potential. We find relations of the parameters of the potentials given in Eq.~(\ref{Eq:lambda}) in terms of physical scalar masses and the angles $\{\beta, \alpha\}$,  then use those as the input parameters for further analysis.

The right-handed up/down quarks and lepton fields are also charged under the aforementioned $Z_{2}$ symmetry and transform as $u^{i}_{R} \to -u^{i}_{R}$, $d^{i}_{R} \to -d^{i}_{R}$ and $\ell^{i}_{R} \to -\ell^{i}_{R}$, respectively. From these charge assignments one realises that all charged fermions exclusively couple to the $\Phi_{2}$ field, leading to the traditional 2HDM Type-I scenario. In Eq.~(\ref{eq:yukeq}) we write down the Yukawa part of the Lagrangian in the mass eigenstate basis.  

\begin{equation}
		\label{eq:yukeq}
  \begin{split}
      - \mathcal{L}_{Yukawa} & = +\sum_{f = u,d,\ell} \left[ m_{f}f\bar{f} + \left(\frac{m_{f}}{v}\kappa^{f}_{h}\bar{f}fh + \frac{m_{f}}{v}\kappa^{f}_{H}\bar{f}fH - i\frac{m_{f}}{v}\kappa^{f}_{A}\bar{f}\gamma_{5}fA\right)\right] \\
             & ~ + \frac{\sqrt{2}}{v}\bar{u}\left(m_{u}V\kappa^{u}_{H^{+}}P_{L} + Vm_{d}\kappa^{d}_{H^{+}P_{R}}\right)dH^{+} + \frac{\sqrt{2}m_{\ell}\kappa^{\ell}_{H^{+}}}{v}\bar{\nu}_{L}\ell_{R}H^{+} + h.c.
  \end{split}	
\end{equation}
Here $m_{f}$ is the fermion mass, $V$ is the Cabibbo-Kobayashi-Maskawa (CKM) matrix and $P_{L/R} = \frac{1 \pm \gamma_{5}}{2}$ are the projection operators. The explicit form of the scaling functions $\kappa_{i}$'s (also called coupling modifiers) are detailed in Table.\ref{tab:couplings}.

\begin{table}[!ht]
    \centering
    \begin{tabular}{l l}
    \toprule[1pt]
        $\kappa^{i}_{S}$ & \ \ \ \ \ \ \ Coefficient \\
        \midrule[1pt]
        $\kappa^{V}_{H}$ & \ \ \ \ \ \ \ $\cos(\beta - \alpha)$ \\
        $\kappa^{V}_{h}$ & \ \ \ \ \ \ \ $\sin(\beta - \alpha)$ \\
        $\kappa^{f}_{H}$ & \ \ \ \ \ \ \ $\frac{\sin \alpha}{\sin \beta}$ \\
        $\kappa^{f}_{h}$ & \ \ \ \ \ \ \ $\frac{\cos \alpha}{\sin \beta}$ \\
        $\kappa^{f}_{A}$ & \ \ \ \ \ \ \ $\cot\beta$ \\
        $\kappa^{u}_{H^{+}}$ & \ \ \ \ \ \ \ $\cot\beta$ \\
        $\kappa^{d/\ell}_{H^{+}}$ & \ \ \ \ \ \ \ $-\cot\beta$ \\
    \bottomrule[1pt]
    \bottomrule[1pt]
    \end{tabular}
    \caption{Explicit form of different coupling modifiers $\kappa^{i}_{S}$. Here $S$ denotes different scalars in the 2HDM and $i$ can be SM gauge bosons and fermions.}
    \label{tab:couplings}
\end{table}

\section{Constraints}
In this section, we describe different theoretical and experimental constraints which are required to restrict the parameter space of the 2HDM Type-I. 

\label{sec:constraints}

\subsection{Theoretical Constraints}
\begin{itemize}
    \item \textbf{Vacuum Stability:} 
    The vacuum stability conditions ensure that the potential must be bounded from below in all possible field direction. To achieve this, the $\lambda_{i}$ parameters must follow certain relationships such that the quartic terms in the potential must dominate for large field values. Here we list down the  conditions on the $\lambda_i$'s which are needed to meet the stability criteria, which prevents the potential from becoming infinitely negative,  \cite{Coleppa_2014}
    \begin{equation*}
    \lambda_1  > 0, \,\,\,\,\, \lambda_2  > 0,  \,\,\,\,\, \lambda_3 + \sqrt{\lambda_1 \lambda_2}  > 0,  \,\,\,\,\, \lambda_3 + \lambda_4 - |\lambda_5| + \sqrt{\lambda_1 \lambda_2}  > 0.
     \end{equation*}
    
    \item \textbf{Unitarity:} The unitarity constraints are necessary to ensure that the theory remains predictive at high energies. At tree level, unitarity imposes specific conditions on the energy growth of all possible $2 \to 2$ scattering processes. Ref.~\cite{Ginzburg_2005,Bhattacharyya:2015nca} derives the unitarity conditions for the 2HDM model explicitly. According to the unitarity constraint, the following relations should be obeyed:
    \begin{equation*}
         |u_i| \leq 8\pi ,
    \end{equation*}
    where
     \begin{equation*}
    \begin{split}
    u_1 & = \frac{1}{2}(\lambda_1 + \lambda_2 \pm \sqrt{(\lambda_1 - \lambda_2)^2 + 4|\lambda_5|^2}), \\ 
    u_2 & = \frac{1}{2}(\lambda_1 + \lambda_2 \pm \sqrt{(\lambda_1 - \lambda_2)^2 + 4\lambda_4^2}), \\
    u_3 & = \frac{1}{2}(3(\lambda_1 + \lambda_2) \pm \sqrt{9(\lambda_1 - \lambda_2)^2 + 4(2\lambda_3+\lambda_4)^2}), \\
    u_4 & = \lambda_3 + 2\lambda_4 \pm 3|\lambda_5|, \\
    u_5 & = \lambda_3 \pm |\lambda_5|, \\
    u_6 & = \lambda_3 \pm \lambda_4. \\
    \end{split}
\end{equation*}
    \item \textbf{Perturbativity:} The perturbativity condition on the parameters of the Higgs potential, which imposes an upper limit on all the quartic couplings, demands that, for all $i$ values, one has $\lambda_i \leq |4\pi|.$
\end{itemize}

\subsection{Experimental Constraints}

\begin{itemize}
    \item \textbf{EW Precision Tests} We evaluated the EW precision constraints by computing the $S, T$ and $U$ parameters using the SPheno package \cite{Porod_2003}, with the model files written in SARAH  \cite{Staub_2015}. These so-called `oblique parameters' provide stringent constraints on new physics masses and relevant couplings. Therefore any BSM theory should conform to these precision data which are primarily collected by LEP, SLC and  Tevatron. The numerical values, with correlation coefficients of $+0.92$ between S and T plus $-0.68$ ($-0.87$) between S and U (T and U) are \cite{2018}
\begin{equation*}
    S = 0.04 \pm 0.11, \,\,\,\,\,\,\, T = 0.09 \pm 0.14, \,\,\,\,\,\,\, U  = -0.02 \pm 0.11.         
\end{equation*}

    \item \textbf{BSM Higgs Boson Exclusions} We assessed the exclusion limits from direct searches for  BSM (pseudo)scalars states at the LEP, Tevatron and the LHC. These exclusion limits were evaluated at the $95 \%$ Confidence Level (C.L.) using the HiggsBounds-6 \cite{Bechtle_2020} module via the HiggsTools \cite{Bahl_2023} package. In our analysis we have also demanded that our lighter Higgs must comply with the results of ~\cite{CMS:2024ulc}, where the Higgs particles are produced in association with a massive vector boson or a top anti-quark pair and further decays via leptonic modes.

    \item \textbf{SM-Like Higgs Boson Discovery} We examined the compatibility of our 125 GeV Higgs boson with the discovered SM-like Higgs boson using a goodness-of-fit test. Specifically, we calculated the $\chi$-square value with HiggsSignals-3 \cite{Bechtle_2021} via HiggsTools, comparing the predicted signal strengths of our Higgs boson to those observed experimentally. We retained the parameter spaces that satisfies the condition $\chi_{125}^2 < 189.42$ \cite{Benbrik:2024ptw, Belyaev:2023xnv}, corresponding to a $95 \%$ C.L. with $159$ degrees of freedom. The choice of the upper bound on $\chi^{2} < 189.42$, corresponds to a global $\chi^2$ goodness of fit test at $95 \%$ C.L. ensuring that the predicted properties of the heavier CP even Higgs state (with the mass $m_{H} = 125$ GeV) of the underlying 2HDM Type-I closely match those of the Higgs boson observed at the LHC, thereby keeping the model compatible under current experimental constraints.

    \item \textbf{Flavour Physics} We incorporated constraints from $B$-physics observables, which are sensitive to potential new physics contributions in loop mediated FCNC processes. Specifically, we tested the most stringent bound on the Branching Ratio ($\mathcal{BR}$) of the $B\rightarrow X_s \gamma$ decay using Next-to-Leading Order (NLO) calculations in QCD as discussed in \cite{Borzumati_1998}:
    \begin{equation}
        \mathcal{BR}(B \rightarrow X_s \gamma) = \frac{\Gamma (B \rightarrow X_s \gamma)}{\Gamma_{SL}}\mathcal{BR}_{SL}
    \end{equation}
    where $\mathcal{BR}_{\rm SL}$ is the semi-leptonic ${\cal  BR}$ and $\Gamma_{SL}$ is the semi-leptonic decay width.
    
%%%%%%%%%%%%%%%%%%%%%%%%%%%%%%%%%%%%%%%%%%%%%%%%%%%%%%%
%    The couplings are defined as 
%
%    \begin{equation*}
%        X  = -\cot{\beta} \, \, \, \, \, \, \, \, \, \, \, Y  =  \cot{\beta} 
%    \end{equation*}
%%%%%%%%%%%%%%%%%%%%%%%%%%%%%%%%%%%%%%%%%%%%%%%%%%%%%%%

    We took our input parameters from the most recent Particle Data Group (PDG) compilation \cite{Workman:2022ynf}, as follows:
    \begin{align*}
         \alpha_s(M_Z) & = 0.1179 \pm 0.0010 , \ \ \ \ \ \ \ \ \ \ \ \ \ \ m_t = 172.76 \pm 0.3, \\
         \frac{m_b}{m_c} & = 4.58 \pm 0.01, \ \ \ \ \ \ \ \ \ \ \ \ \ \ \alpha  = \frac{1}{137.036}, \\
        \cal BR_{\rm SL} & = 0.1049 \pm 0.0046, \ \ \ \  |\frac{V_{ts}^*V_{tb}}{V_{cb}}|^2 = 0.95 \pm 0.02, \\
         m_b(\overline{\rm MS}) & = 4.18 \pm 0.03, \ \ \ \ \ \ \ \ \ \ \ \ \ \ m_c = 1.27 \pm 0.02, \\
        m_Z & = 91.1876 \pm 0.0021, \ \ \ \ \ \ \ \ \  m_W = 86.377 \pm 0.012.
    \end{align*}
    %This analysis is critical for ruling out parameter space regions that would lead to significant discrepancies in flavor physics. 
    The following restriction has been imposed, which represents the $3 \sigma$ experimental limit: 
    \begin{equation*}
         2.87 \times 10^{-4} < \mathcal{BR}(B \rightarrow X_s \gamma) < 3.77 \times 10^{-4}.
    \end{equation*}
    Other $B$-physics observables, like $\mathcal{BR}(B^+ \rightarrow \tau^+ \nu_\tau)$, $\mathcal{BR}(D_s \rightarrow \tau \nu_\tau)$, $\mathcal{BR}(B_s \rightarrow \mu^+ \mu^-)$ and  $\mathcal{BR}(B^0 \rightarrow \mu^+ \mu^-)$ have been taken care of by using the FlavorKit tool \cite{Porod_2014} provided by the SPheno package  \cite{Porod_2003}. Our calculated $b \rightarrow s \gamma$ results were also found to be consistent with the FlavorKit tool.
\end{itemize}

\section{Explaining the anomalies}
\label{sec:anomaly}

The primary objective of this paper is to investigate whether the 2HDM Type-I can explain the excesses observed at the LHC and in LEP data over the $94-96$ GeV  mass range. To do so, we need to define the signal strength corresponding to these three excesses. The signal strength is formulated as a ratio between the observed number of events to the expected number of events for a hypothetical SM Higgs state of mass $95$ GeV. Assuming the Narrow Width Approximation (NWA), the signal strength for the $\tau^+ \tau^-, \gamma \gamma$ and $b \Bar{b}$ channels can be parameterised as cross section ($\sigma$) times $\mathcal{BR}$,
as follows:
\begin{equation}
    \begin{split}
        \mu_{\tau^+ \tau^-}(\phi) & = \frac{\sigma_{\rm 2HDM}(gg\rightarrow \phi)}{\sigma_{\rm SM}(gg\rightarrow h_{\rm SM})} \times \frac{\mathcal{BR}_{\rm 2HDM}(\phi \rightarrow \tau^+ \tau^-)}{\mathcal{BR}_{\rm SM}(h_{\rm SM} \rightarrow \tau^+ \tau^-)}, \\
        \mu_{\gamma \gamma}(\phi) & = \frac{\sigma_{\rm 2HDM}(gg\rightarrow \phi)}{\sigma_{\rm SM}(gg\rightarrow h_{\rm SM})} \times \frac{\mathcal{BR}_{\rm 2HDM}(\phi \rightarrow \gamma \gamma)}{\mathcal{BR}_{\rm SM}(h_{\rm SM} \rightarrow \gamma \gamma)}, \\
        \mu_{b \Bar{b}}(\phi) & = \frac{\sigma_{\rm 2HDM}(e^+e^-\rightarrow Z \phi)}{\sigma_{\rm SM}(e^+e^-\rightarrow Z h_{\rm SM})} \times \frac{\mathcal{BR}_{\rm 2HDM}(\phi \rightarrow b \Bar{b})}{\mathcal{BR}_{\rm SM}(h_{\rm SM} \rightarrow b \Bar{b})}.
    \end{split}
    \label{Eq:sigstrength}
\end{equation}
Here, $h_{\rm SM}$ corresponds to a SM like Higgs Boson with a mass of 95 GeV while $\phi$ is a 2HDM Type-I Higgs state with the same mass. The experimental measurements for these three signal strengths around $95$ GeV are expressed as 
\begin{equation}
    \begin{split}
        \mu_{\gamma \gamma}^{\rm exp} & = \mu_{\gamma \gamma}^{\rm ATLAS+CMS} = 0.24^{+0.09}_{-0.08}, ~\text{\cite{CMS-PAS-HIG-20-002, CMS:2018cyk, ATLAS-CONF-2023-035}}   \\
        \mu_{\tau^+ \tau^-}^{\rm exp} & = 1.2 \pm 0.5, ~\text{\cite{CMS:2022goy}} \\
        \mu_{b \Bar{b}}^{\rm exp} & = 0.117 \pm 0.057. ~\text{\cite{LEPWorkingGroupforHiggsbosonsearches:2003ing, Cao_2017}}
    \end{split}
    \label{eq:signal_excess}
\end{equation}

Although the ditau excess is most prominent around $100$ GeV and the $b \Bar{b}$ excess near $98$ GeV, a search around $95$ GeV could provide a unified explanation for all these three anomalies. This is because the mass resolution in the ditau final states is substantially large and the LEP excess, associated with the $b \bar{b}$ final states, is also rather broad: therefore, a common origin for these excesses may plausibly reside around $95$ GeV \cite{Azevedo:2023zkg}.

In our analysis, we have combined the di-photon measurements from the ATLAS and CMS experiments, denoted as $\mu_{\gamma \gamma}^{\rm ATLAS}$ and $\mu_{\gamma \gamma}^{\rm CMS}$, respectively. The ATLAS measurement yields a central value of $0.18 \pm 0.1$  \cite{PhysRevD.109.035005} while the CMS measurement yields a central value of $0.33^{+0.19}_{-0.12}$  \cite{Biek_tter_2023}. The combined measurement, denoted by $\mu_{\gamma \gamma}^{\rm ATLAS+CMS}$ is determined by taking the average of these two central values, assuming  these measurements are uncorrelated. The corresponding combined uncertainty is calculated by adding the individual uncertainties in quadrature.  To determine whether the observed excess can be explained through our model or otherwise, we perform a $\chi^2$ analysis using the central values $\mu^{\rm exp}$ and the $1 \sigma$ uncertainties $\Delta \mu^{\rm exp}$ associated with the signal related to the excesses as defined in Eq. (\ref{eq:signal_excess}). The contribution to the $\chi^2$ for each of the channel is calculated using the equation
\begin{equation}
    \chi^2_{\gamma \gamma, \tau^+ \tau^-, b \Bar{b}} = \frac{(\mu_{\gamma \gamma, \tau^+ \tau^-, b \Bar{b}}(\phi) - \mu_{\gamma \gamma, \tau^+ \tau^-, b \Bar{b}}^{\rm exp})^2}{(\Delta \mu_{\gamma \gamma, \tau^+ \tau^-, b \Bar{b}}^{\rm exp})^2}.
\end{equation}
Hence, the resulting $\chi^2$, which we will use to determine if the excesses can be achieved by the allowed region of the parameter space of 2HDM Type-I, or otherwise, is the following:
\begin{equation}
    \label{eq:chisum}
    \chi^2_{\gamma \gamma, \tau^+ \tau^-, b \Bar{b}} = \chi^2_{\gamma \gamma} + \chi^2_{\ \tau^+ \tau^-} + \chi^2_{b \Bar{b}}.
\end{equation}

We test this BSM scenario for two cases: firstly, we consider both the CP-even and CP-odd  Higgs states simultaneously (i.e., $\phi=h+A$, except for $b\bar b$ where $\phi=h$ as the CP-odd scalar $A$ does not couple to $ZZ$ mode) in explaining the anomaly  and, secondly, we only exploit the CP-even Higgs state (i.e., $\phi=h$) in order to explain it. Hence, we align our $H$ state (recall that we have $m_h<m_H$) with the SM Higgs boson, so that  $m_H=125$ GeV,  and start a Monte Carlo (MC) sampling of the various input parameters.

\subsection{The Overlapping Solution}
In the case of the overlapping solution, the signal strengths corresponding to the $\tau^+ \tau^-$ and $\gamma \gamma$ channels receive substantial contribution from both the CP-even and CP-odd states simultaneously\footnote{Here we have considered a CP-conserving potential. As a result, the $h$ and $A$ states would not interfere with each other.}. In contrast, for the $b \Bar{b}$ mode only the CP-even state contribute as the trilinear coupling $AZZ$ is zero at tree level. As a result, the signal strengths can be expressed in the following manner:
\begin{equation*}
    %\begin{split}
        \mu_{\tau^+ \tau^-}(h+A)  = \mu_{\tau^+ \tau^-}(h) + \mu_{\tau^+ \tau^-}(A),~~~ 
        \mu_{\gamma \gamma}(h+A)  = \mu_{\gamma \gamma}(h) + \mu_{\gamma \gamma}(A),~~~
        \mu_{b\bar b} (h).  ~~~
   % \end{split}
\end{equation*}
\begin{table}[!t]
    \centering
    \begin{tabular}{l l}
    \toprule[1pt]
        Parameter & \ \ \ \ \ \ \ Scan Range \\
        \midrule[1pt]
        $m_H$ & \ \ \ \ \ \ \ $125$ GeV \\
        $m_h$ & \ \ \ \ \ \ \  $94~\text{GeV}-96~\text{GeV}$ \\
        $m_A$ & \ \ \ \ \ \ \ $94~\text{GeV}-96~\text{GeV}$  \\
        $m_{H^\pm}$ & \ \ \ \ \ \ \ $140~\text{GeV}-250~\text{GeV}$  \\
        $\tan{\beta}$ & \ \ \ \ \ \ \ $0.5-100$  \\
        $\sin({\beta - \alpha})$ & \ \ \ \ \ \ \ $-0.60 - 0.60$ \\
    \bottomrule[1pt]
    \bottomrule[1pt]
    \end{tabular}
    \caption{The scan range which is used for the MC sampling for the overlapping solution.}
    \label{tab:scanrange}
\end{table}
\begin{figure}[!t]
    \centering
    \includegraphics[scale=0.4]{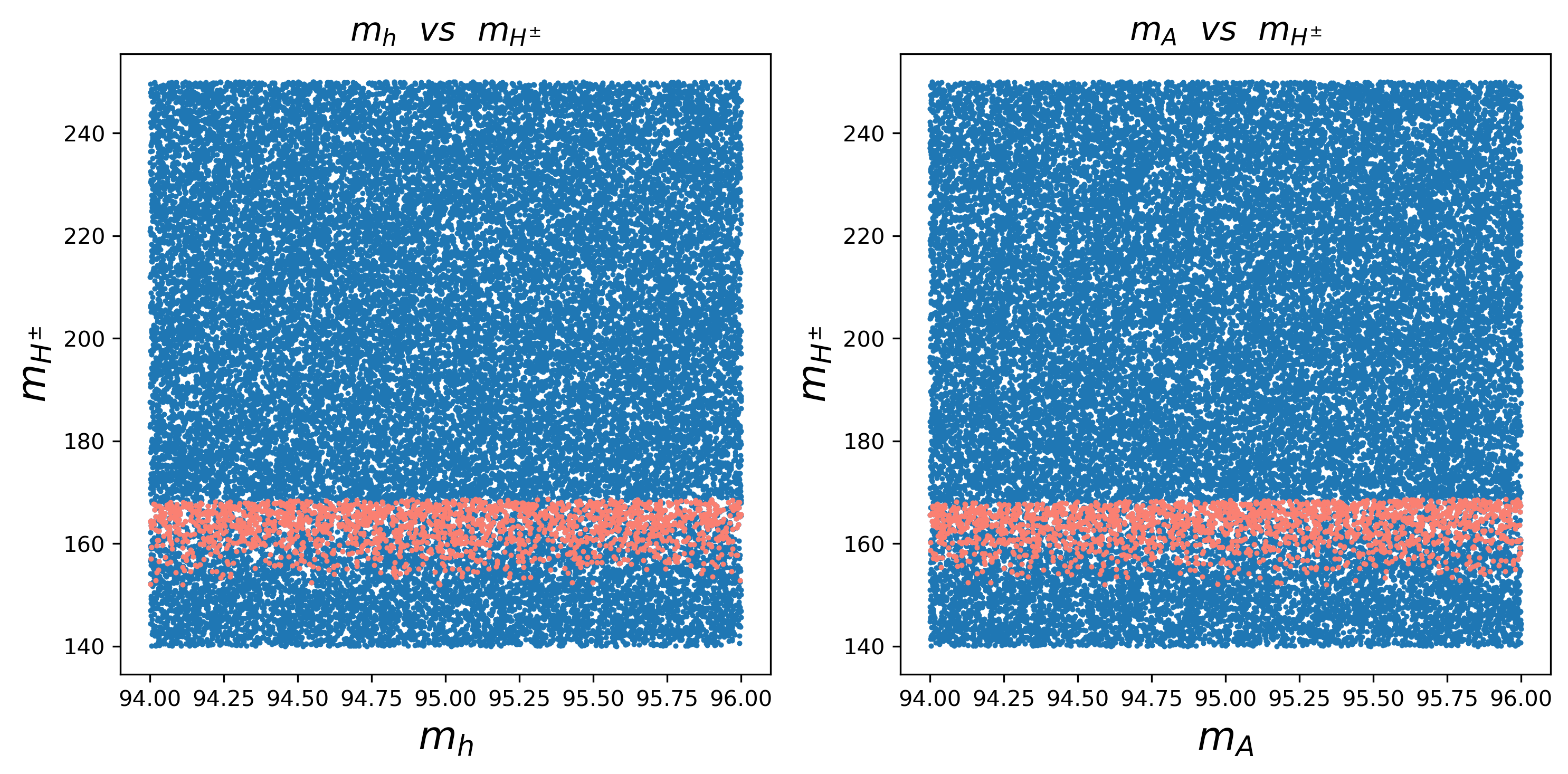}
    \caption{Results of the scan in Table~\ref{tab:scanrange} mapped against the Higgs boson masses. The blue regions are allowed by stability, unitarity and perturbativity constraints while the red  regions are  allowed by Higgs data, $b \rightarrow s \gamma $ and  EW precision constraints.}
    \label{fig:simultaneous_param}
\end{figure}

We generated MC samples in the scan range as described in Table ~\ref{tab:scanrange}. After testing them against various theoretical and experimental constraints, the allowed parameter space is illustrated in Figure~\ref{fig:simultaneous_param}. The region of the parameter space that pass the theoretical constraints are depicted by the blue points while the region that pass the experimental constraints are depicted by red points. The plot illustrates that a nearly degenerate solution with both the CP-even and CP-odd Higgs in the $94-96$ GeV mass range is viable for the charged Higgs mass ranging over the interval $152 ~\text{GeV} < m_H^\pm < 168 ~\text{GeV}$. We will use the overlapping region of the two coloured point distributions to test the aforementioned anomalies.

\begin{figure}[h!]
    \centering
    \includegraphics[width=\linewidth]{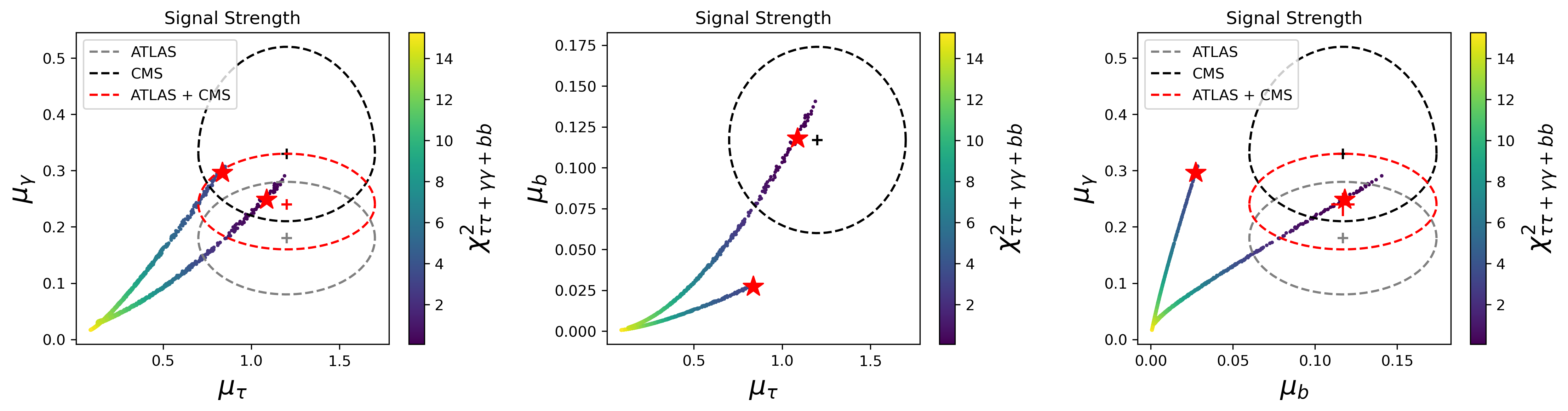}
    \caption{Correlations amongst the signal strengths in Eq. (\ref{Eq:sigstrength}). Here, the total $\chi^2$ is displayed using the colour bar and the best fit point is given by the star marker. The ATLAS and CMS results with their corresponding $1 \sigma$ band are also represented.}
    \label{fig:simultaneous_chi}
\end{figure}

\begin{figure}[!t]
    \centering
    \includegraphics[scale=0.35]{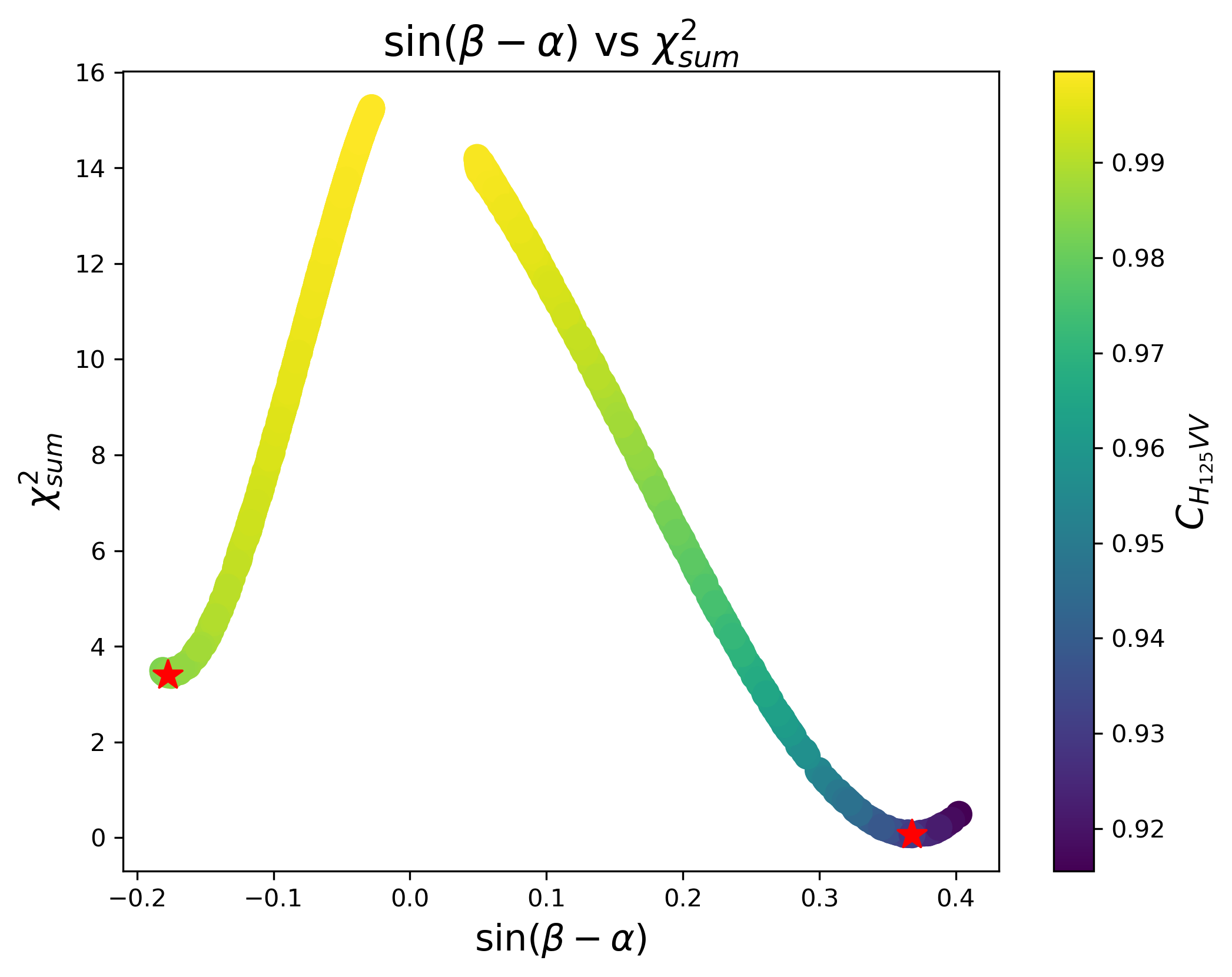}
    \caption{The value of $\sin(\beta-\alpha)$ plotted against $\chi^2_{\rm sum}$. The $\chi^2_{\rm min}$ is indicated by a star. The color gradient represents the strength of the coupling between the SM like Higgs boson and the pair of vector bosons. }
    \label{fig:simul_sin}
\end{figure}

Figure \ref{fig:simultaneous_chi} illustrates the total $\chi^2$ distribution for points that are compatible with all three anomalies and the best fit point is marked by a star, corresponding to $\chi^2_{\rm min}$. The experimentally observed signal strength with their $1 \sigma$ band is also superimposed in the plot to test them against this model's best fit point. The chi-square fit reveals two distinct branches, differentiated by the sign of $\sin(\beta - \alpha)$. The longer branch is characterised by the positive values, while the smaller one by  the negative values.  Figure \ref{fig:simul_sin} further highlights the sign of the EW coupling parameter, with the best-fit point in both  branches again marked by a red star. Notably, the positive branch contains a best fit point that simultaneously explains all the three anomalies within the $1 \sigma$ region. However, an analysis by the ATLAS collaboration on the 2HDM Type-I~\cite{ATLAS:2021vrm} imposes significant restrictions on the positive region of the value $\sin(\beta - \alpha)$, limiting its viability. The negative branch also allows for a simultaneous explanation of all the three anomalies within the $1 \sigma$ region. However, when the $b \Bar{b}$ channel anomaly is examined in isolation, it cannot be explained at the $1 \sigma$ level. While the vector boson coupling with the SM Higgs boson, indicated by the colour bar, is weakened as in this case, it remains within the allowed range. 

\begin{figure}[h!]
    \centering
    \includegraphics[scale=0.285]{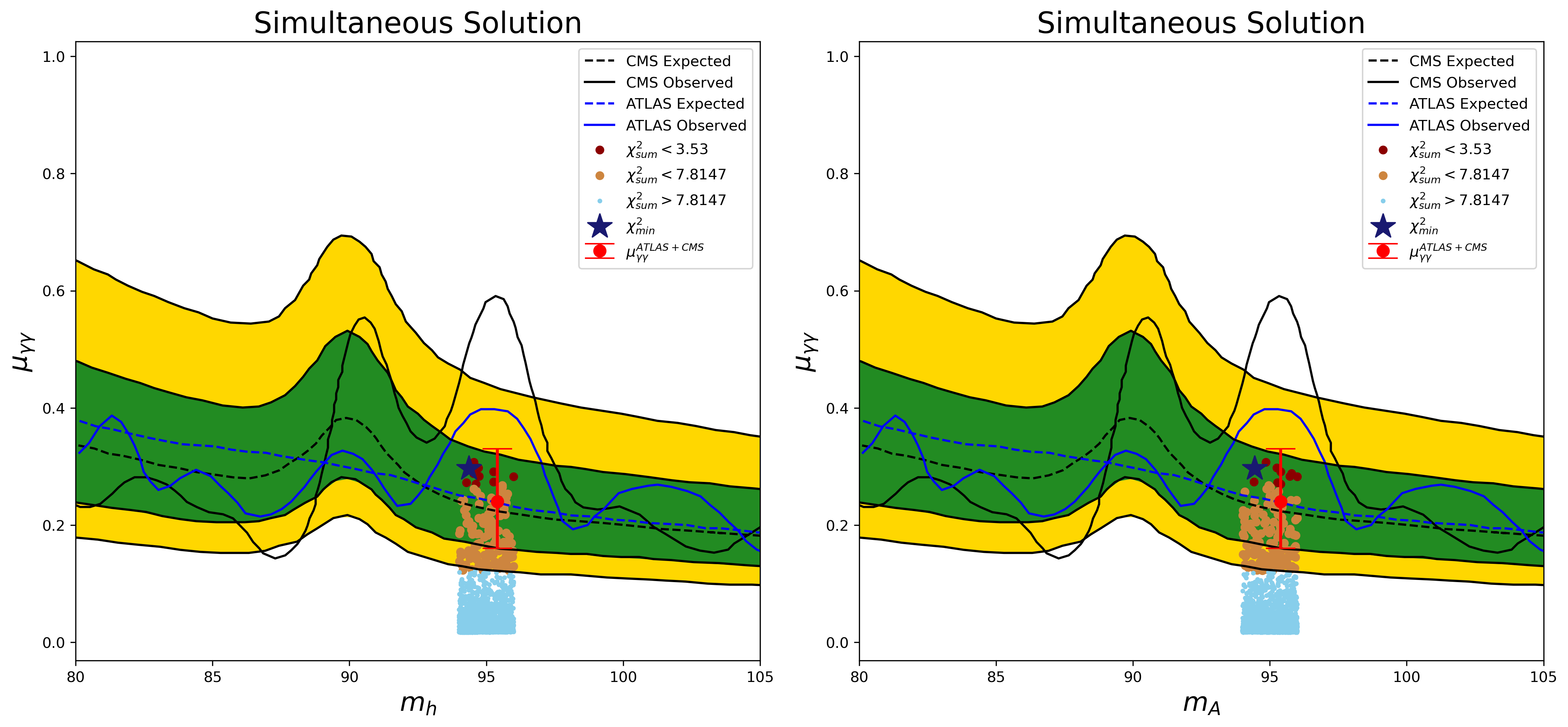}
    \caption{The di-photon signal strength results from experiment tensioned against the 2HDM Type-I predictions satisfying the three anomalies simultaneously.}
    \label{fig:simultaneous_gamma}
\end{figure}

Given that the di-photon excess is most pronounced around $95$ GeV, we also plot the allowed points within the $94-96$ GeV range over the CMS and ATLAS results for the signal strength in the ${\gamma \gamma}$ channel, for the negative branch, as shown in Figure ~\ref{fig:simultaneous_gamma}. The expected and observed CMS limits are shown by the black dashed and solid lines. The green and yellow bands indicate the $1 \sigma$ and $2 \sigma$ uncertainties and the plot is overlaid by the ATLAS observed $95 \%$ C.L. limits on the signal strength with the red dashed and solid lines. The combined signal strength (CMS+ATLAS) at $95.4$ GeV with its error bar is also shown using a red dot. The points explaining the anomaly at the $1 \sigma$ level, for $3$ degrees of freedom, corresponding to the $\gamma \gamma$, $\tau \tau$ and $b \Bar{b}$ channels as shown in Eq.~(\ref{eq:chisum}), which requires $\chi^2 < 3.53$, are plotted in dark red while the points explaining it at the $2 \sigma$ level, demanding $\chi^2 < 7.8147$, are shown in peru colour. (Less likely points are given in sky blue.) The best fit point in the $94-96$ range is also indicated using midnight blue color. It can be clearly seen that many parameter points are suited to explain the excesses observed. The details of the best fit point as marked in the figure is shown in Table~\ref{tab:simul_bench}.
Although we have imposed a global $\chi^2$ goodness of test at $95 \%$ C.L., the best-fit BP listed in Table ~\ref{tab:simul_bench} yields a $\chi^2$ value of $171.14$, , which also falls within the $3 \sigma$ region corresponding to the more stringent HiggsSignals constraint, $\chi^2\leq\chi^2_{\rm SM}+11.83$. 

\begin{table}[!b]
    \centering
        \hspace*{-0.75 truecm}
    \begin{tabular}{l l l l l l l l l l l l l}
    \toprule[1pt]
        Parameter & \ \ \ \ $m_H$ & \ \ \ \ $m_h$ & \ \ \ \ $m_A$ & \ \ \ \ $m_{H^\pm}$  & \ \ \ \ $\tan \beta$ & \ \ \ \ $\sin(\beta - \alpha)$ & \ \ \ \ $\mu_{\tau^+ \tau^-}(h+A)$ & \ \ \ \ $\mu_{\gamma \gamma}(h+A)$ & \ \ \ \ $\mu_{b \Bar{b}}(h)$ & \ \ \ \ $\chi^2_{\tau^+ \tau^- + \gamma \gamma + b \bar{b}}$ \\
        \midrule[1pt]  \\
        Figure~\ref{fig:simultaneous_chi} & \ \ \ \ 125.0 & \ \ \ \ 94.36 & \ \ \ \ 94.44 & \ \ \ \ 166.63 & \ \ \ \ 2.33 & \ \ \ \ $-0.17$ & \ \ \ \ 0.836 & \ \ \ \ 0.296 & \ \ \ \ 0.027 & \ \ \ \ 3.402 \\[0.25cm]
    \bottomrule[1pt]
    \bottomrule[1pt]
    \end{tabular}
    \caption{BP extracted from Figure~\ref{fig:simultaneous_chi} corresponding to the negative region of $\sin(\beta - \alpha)$. The masses of the Higgs states are given in units of GeV. The positive region of $\sin(\beta - \alpha)$ is strongly constrained by the ATLAS analysis of the 2HDM Type-I ~\cite{ATLAS:2021vrm}, rendering it less viable.}
    \label{tab:simul_bench}
\end{table}
 
\subsection{The Single Solution}

%In this case, only the $h$ state is responsible for explaining all the three  anomalies.
%and the ranges scanned for the 2HDM Type-I input parameters are given in Table~\ref{tab:scanrangeeven}.

\begin{table}[h!]
    \centering
    \begin{tabular}{l l}
    \toprule[1pt]
        Parameter & \ \ \ \ \ \ \ Scan Range \\
        \midrule[1pt]
        $m_H$ & \ \ \ \ \ \ \ $125$ GeV \\
        $m_h$ & \ \ \ \ \ \ \  $94~\text{GeV}-96~\text{GeV}$ \\
        $m_A$ & \ \ \ \ \ \ \ $90~\text{GeV}-250~\text{GeV}$  \\
        $m_{H^\pm}$ & \ \ \ \ \ \ \ $140~\text{GeV}-250~\text{GeV}$  \\
        $\tan{\beta}$ & \ \ \ \ \ \ \ $0.5-100$  \\
        $\sin({\beta - \alpha})$ & \ \ \ \ \ \ \ $-0.60 - 0.60$ \\
    \bottomrule[1pt]
    \bottomrule[1pt]
    \end{tabular}
    \caption{The scan range which is used for the MC sampling for the single solution.}
    \label{tab:scanrangeeven}
\end{table}

\begin{figure}[h!]
    \centering
    \includegraphics[width=\linewidth]{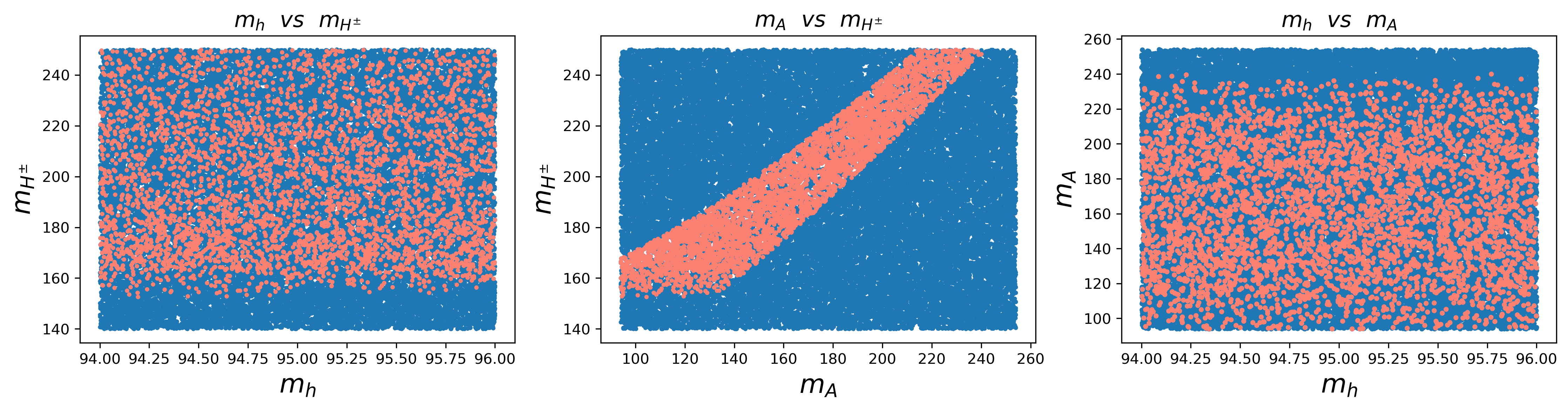}
    \caption{Results of the scan in Table~\ref{tab:scanrangeeven}  mapped against the Higgs masses. The blue regions are allowed by stability, unitarity and perturbativity constraints while the red  regions are  allowed by Higgs data, $b \rightarrow s \gamma $ and  EW precision constraints.}
    \label{fig:single_param}
\end{figure}

In this case, only the $h$ state is responsible for explaining all the three  anomalies. We sampled points for the scan ranges described in Table ~\ref{tab:scanrangeeven} and the points that pass the various constraints are depicted in  Figure~\ref{fig:single_param}, wherein the blue shaded region indicates the points that pass the various theoretical constraints while the red shaded region indicates the part of parameter space that survives after imposing different experimental bounds. The plots clearly represent the allowed parameter space for the CP-odd and charged Higgs masses, given that we fix our CP-even mass to lie in the range $94-96$ GeV. Although the charged Higgs mass is bounded at around $155$ GeV, the CP-odd one covers almost the entire scan range. The plots also clearly illustrate the degeneracy between the CP-odd and the charged Higgs states, which is required to satisfy the EW precision constraints. Note that the allowed CP-odd scalar mass also lies within the $94~\text{GeV}~\text{to}~96$ GeV mass window, hinting to the overlapping solution that we discussed in the previous section. We move ahead with testing the allowed points with the observed anomalies. 

\begin{figure}[h!]
    \centering
    \includegraphics[scale=0.5]{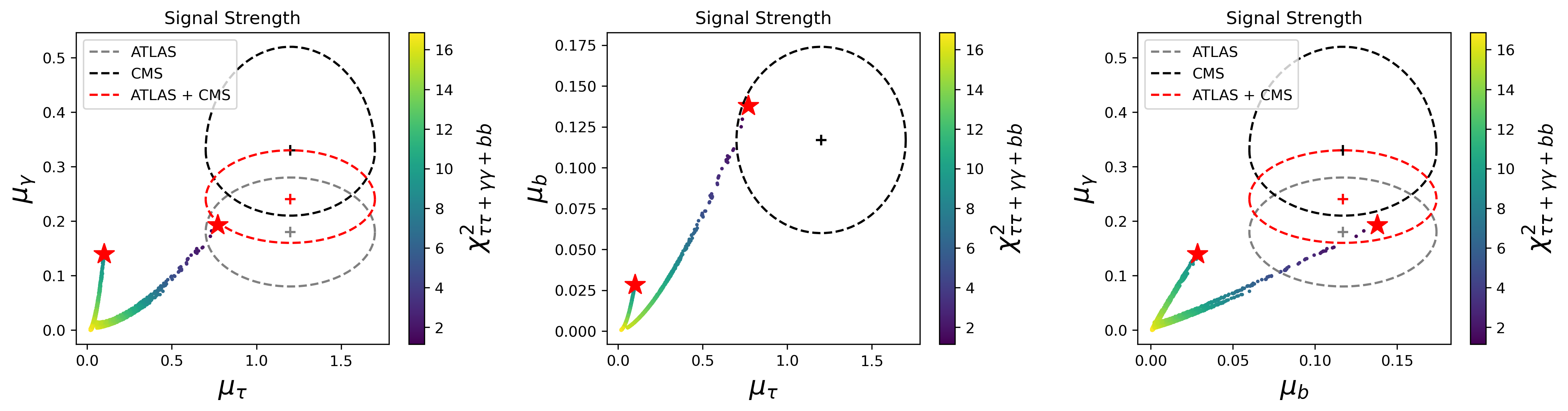}
    \caption{Correlations amongst the signal strengths in Eq. (\ref{Eq:sigstrength}). Here, the total $\chi^2$ is displayed using the colour bar and the best fit point is given by star marker. ATLAS and CMS results with their corresponding $1 \sigma$ bands are also added.}
    \label{fig:single_chi}
\end{figure}

\begin{figure}[!tbp]
  \centering
    \includegraphics[scale=0.45]{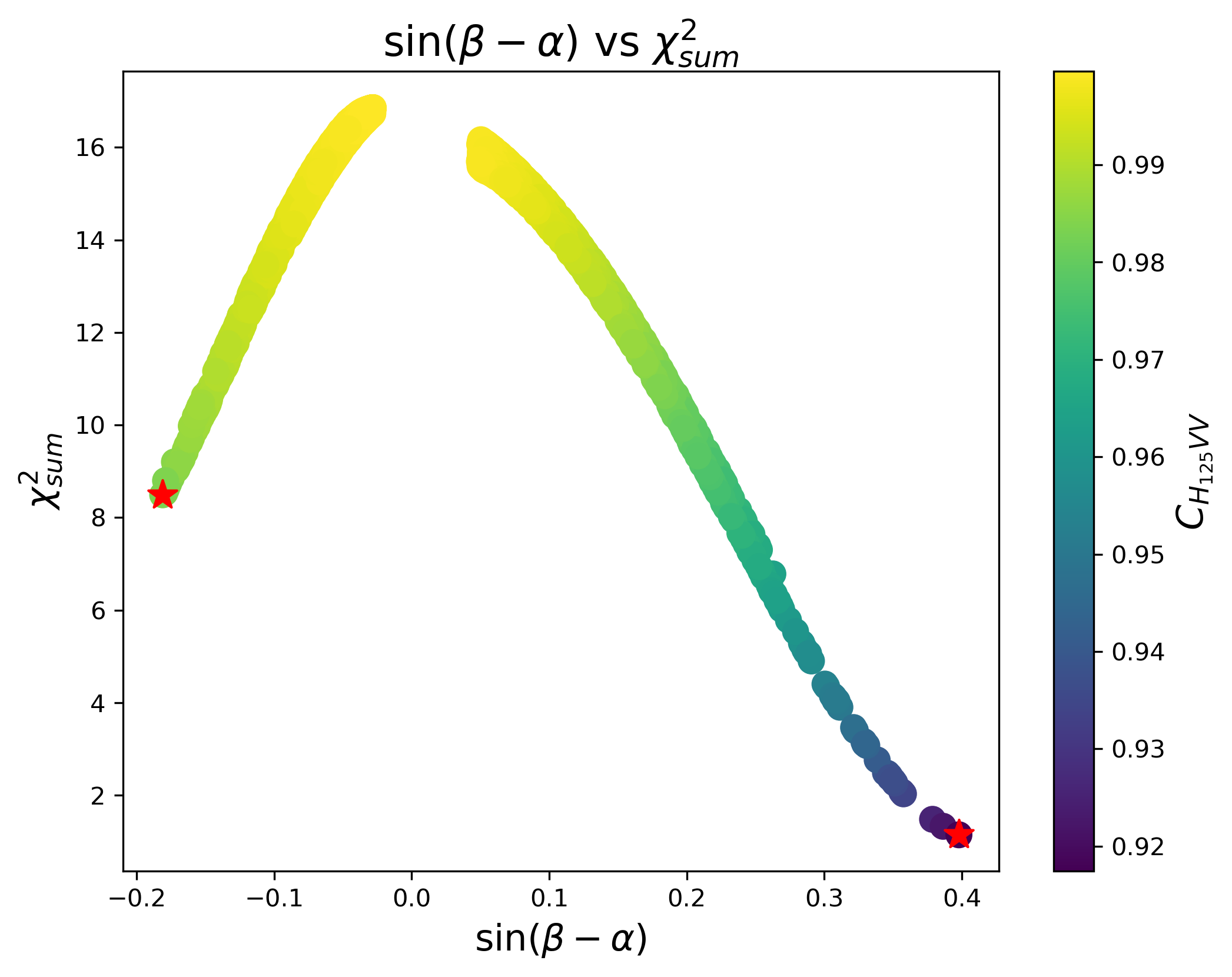}
    \caption{The value   of $\sin(\beta-\alpha)$  plotted against $\chi^2_{\rm sum}$. The best fit point is indicated by a star. The color gradient represents the strength of the coupling between the SM like 125 GeV Higgs boson and the pair of vector bosons.}
    \label{fig:single_sin}
\end{figure}

The total $\chi^2$ fit for the points passing all the constraints is displayed in Figure~\ref{fig:single_chi}, wherein the best fit point (i.e., again, the $\chi^2_{\rm min}$ one) is indicated with a star. We have again overlaid the plot with the experimentally observed data within the $1 \sigma$ band. Here again the branches are differentiated by the sign of $\sin(\beta - \alpha)$, with the larger branch depicting the positive while the smaller one depicting the negative values. The plot demonstrates that, while a simultaneous solution exists for all three cases in the positive branch, no point in the negative branch satisfies all three anomalies simultaneously within the $2\sigma$ level ($\chi^2 < 7.8147$), once the ATLAS constraint is imposed. Nevertheless, viable points do appear in the negative branch when the requirement is relaxed to the $3\sigma$ level ($\chi^2 < 14.156$). Figure \ref{fig:single_sin} provides additional insight into the sign of the electroweak coupling parameter, with red stars denoting the best-fit points in each of the two branches.

\section{Conclusions}
\label{sec:conclusions}
In summary, we have proven that somewhat anomalous data produced at LEP and the LHC in $b\bar b$ as well as $\tau^+\tau^-$ and $\gamma\gamma$ final states, respectively, all clustering in a mass window around 95 GeV, are consistent with the possibility of the 2HDM Type-I in inverted mass hierarchy (i.e., $m_H=125$ GeV) explaining these through one of its neutral Higgs states. Specifically, only one configuration is possible: the one where both the (degenerate) $h$ and $A$ states cooperate to explain the aforementioned anomalies, if we take the ATLAS result into account. This is an intriguing finding, as such a Higgs state can also be probed in collateral signatures specific to the 2HDM Type-I in inverted mass hierarchy, as emphasised in various previous literature
\cite{Arhrib:2016wpw,Arhrib:2017wmo,Arhrib:2017uon,Arhrib:2020tqk,Arhrib:2021xmc,Wang:2021pxc,Arhrib:2021yqf,Li:2023btx}. To aid testing this theoretical hypothesis, we have produced a BP amenable to further phenomenological analysis, wherein the parameter space point is giving the best fit to all anomalies. We also found that there does not exist a solution for only the CP-even case, that could explain the three anomalies simultaneously in the 2HDM Type-I even within $2 \sigma$.

\section*{Acknowledgments}
The work of S.M. is supported in part through the NExT Institute and the STFC Consolidated Grant  ST/X000583/1. A.S. acknowledges the support from SERB-National Postdoctoral fellowship (Ref No: PDF/2023/002572). A.K. acknowledges the support from Director's Fellowship at IIT Gandhinagar. All authors thank Tanmoy Mondal and Prasenjit Sanyal for their help in discovering a mistake in their calculations.

%\appendix
%\bibliographystyle{apsrev4-2}
\bibliography{bibliography}

\end{document}